\documentstyle[preprint,float,floats,pra,aps,epsfig]{revtex}
\tightenlines  
\newcommand{\Lp}{L_x+\text{i}\eta L_y}
\newcommand{\sqe}{\sqrt{1-\eta^2}}
\newcommand{\mylims}{\stackrel{\scriptscriptstyle N\gg 1}
{\longrightarrow}}

\begin{document} \draft \textheight=9in \textwidth=6.5in 

\title{\LARGE \bf Squeezed angular momentum coherent states~$\!$:\protect\\
 construction and time evolution\footnote{presented on ICSSUR '99:
 6th International Conference on Squeezed States and Uncertainty 
 Relations, Napoli, 24-29 May 1999
}}

\author{\underline{R. ARVIEU}}
\address{Institut des Sciences Nucl\'eaires - F 38026 - Grenoble C\'edex -
France\\ e-mail address: arvieu@isn.in2p3.fr}
\author{P. ROZMEJ}
\address{University Marie Curie Sklodowska - PL 20-031 - Lublin - Poland\\
e-mail address: rozmej@tytan.umcs.lublin.pl}
%

\maketitle

\begin{abstract}
 A family of angular momentum coherent states on the sphere 
is constructed
using previous work by Aragone et al [1]. These states depend on a complex
parameter which allows an arbitrary squeezing of the angular momentum
uncertainties.  The time evolution of these states is analyzed assuming a
rigid body hamiltonian.  The rich scenario of fractional revivals is
exhibited with cloning and many interference effects. 
\end{abstract}

\vspace{8mm}

     In this contribution we will concentrate on a family of coherent
states on the sphere which can be proposed for the description of the
rotation of quantum simple systems like rigid diatomic molecules or rigid
nuclei. The relevant hamiltonian depends then only on the angular momentum
$I$ and the energy spectrum is expressed in terms of a frequency 
$\omega_0$ by $E_I=\hbar \omega_0 \,I(I+1)$. 
A general wave packet (WP) of the family depends on
a parameter $\eta$ and of a real integer number $k$ and will be denoted as 
$\Psi_{\eta k} (\theta,\phi)$. 
The states with $k$ different from 0 are deduced from a
parent state $\Psi_{\eta 0}$ defined as  
\begin{equation}\label{lab1} 
\Psi_{\eta\, 0}(\theta,\phi) 
= \sqrt{\frac{N}{2\pi\sinh 2N}} e^{N\sin\theta(\cos\phi+\text{i}\eta\sin\phi)} 
\end{equation}				     
 For real $\eta$ the angular spread of the latter depends only on $N$ while the
average value of $L_z$ is given by 
\begin{equation}\label{lab2} 
\langle L_z\rangle = \eta(N\coth (2N)-\case{1}{2}) \mylims
\eta(N-\case{1}{2})
\end{equation}				     
 The states with $k\neq 0$ are obtained from (\ref{lab1}) by application on 
$\Psi_{\eta\,0}$ of an operator $({{\cal L}_+})^k$. 
The  operator ${\cal L}_+$ and two other ones which form an
SU(2) algebra are defined by
\begin{equation}\label{lab3} 
{\cal L}_3 = \frac{\Lp}{\sqe}, \qquad
{\cal L}_{\pm} = \pm \left(\frac{\eta L_x +\text{i}L_y}{\sqe}\right) - L_z
\end{equation}				     

 Up to a normalization factor we have 
\begin{equation}\label{lab4} 
\Psi_{\eta\,k}(\theta,\phi) = ({{\cal L}_+})^k \, \Psi_{\eta\,0}(\theta,\phi)
\end{equation}

\begin{figure}[H]
   \begin{center}
     \epsfxsize=10.1cm \epsfbox{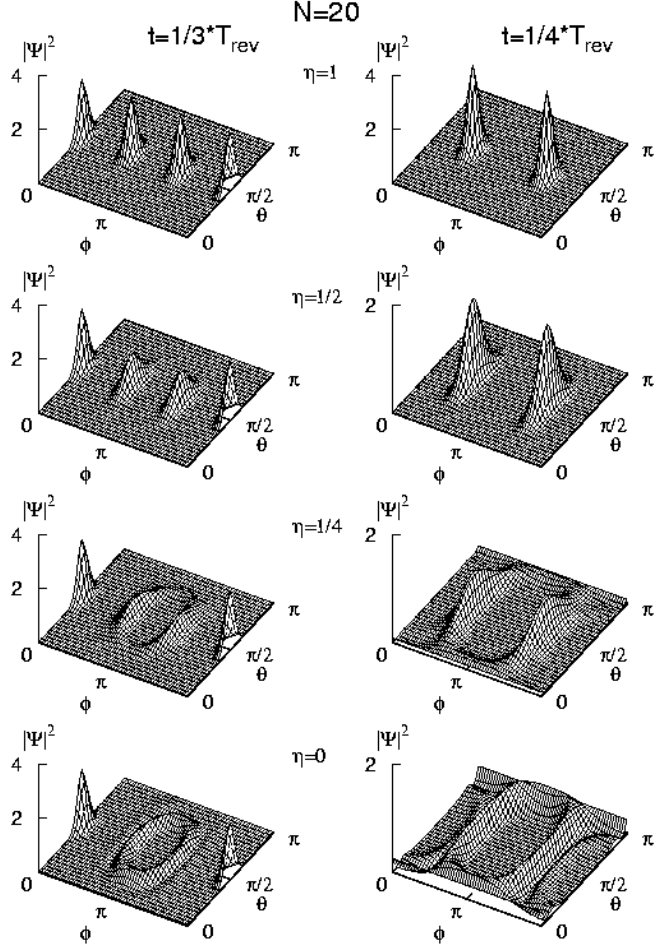}
   \end{center}
\caption[clonring]{Transition of fractional wave packets
from exact clones ($\eta=1$) through developing crescents
($\eta=1/2,\eta=1/4$) to ring topology ($\eta=0$) is demonstrated
for two fractional revival times $t=1/3*T_{\text{rev}}$ (left)
and $t=1/4*T_{\text{rev}}$ (right). The fractional waves called mutants
are clearly seen in the lower rows of the figure.}
\label{ringclon}
\end{figure}
				     
 The states $\Psi_{\eta\, k}$ have the following properties:
 \begin{enumerate}
 \item They are eigenstates of ${\cal L}_3$
\begin{equation}\label{lab5} 
{\cal L}_3 \,\Psi_{\eta\,k} = k\sqrt{1-\eta^2}\, \Psi_{\eta\,k} 
\end{equation}				     
 \item The parameter $\eta=|\eta|\,\exp(\text{i}\alpha)$ is a squeezing parameter 
 since one has  
\begin{equation}\label{lab6} 
|\eta|^2 = \frac{\Delta L_x^2}{\Delta L_y^2}
\end{equation}				     
 \item If $\eta$ is real the WP are minimum uncertainty states and in general we
have
\begin{equation}\label{lab7} 
\Delta L_x^2 \Delta L_y^2 = \frac{1}{4}[\langle L_z\rangle^2 +
|\langle\{ L_x,L_y \}\rangle - \langle L_x\rangle\langle L_y\rangle|^2]
= \frac{1}{4} \frac{\langle L_z\rangle^2}{\cos^2\alpha}
\end{equation}				     
 \item Changing $k$ enable to change the average values of all the components
of L.
\end{enumerate} 

\begin{figure}[H]
  \begin{center}
  \epsfxsize=10.1cm \epsfbox{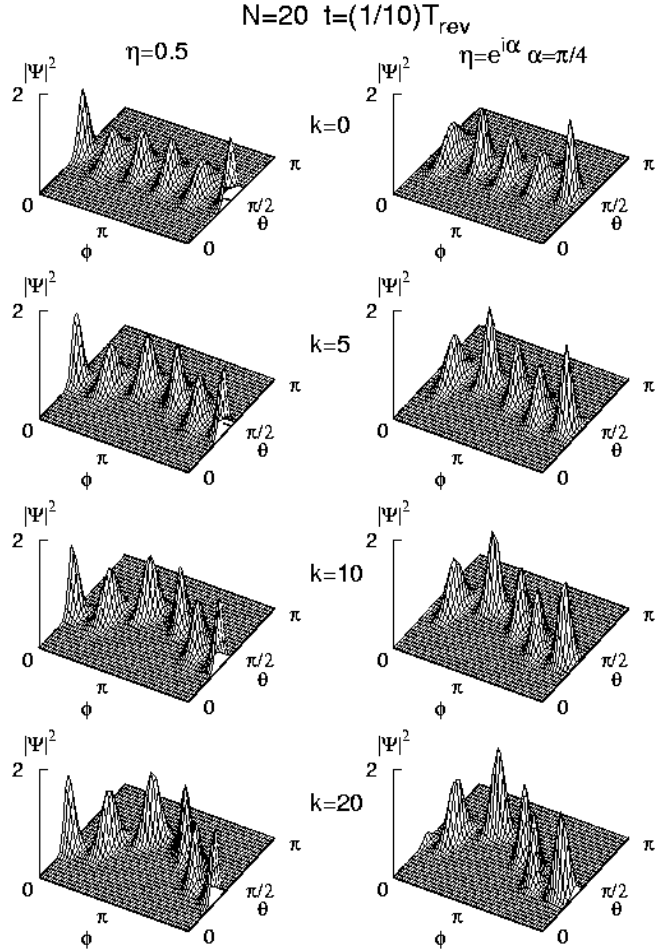} \end{center}
\caption{Shapes of wave packets with $N=20$ at fractional revival
time $t=(1/10)\,T_{\text{rev}}$ for real $\eta=0.5$  (left column), and
$\eta=\exp(\text{i}\alpha),  \alpha=\pi/4$ (right column)
and $k=0,5,10,20$ as funtions of angular variables for a rigid
molecule. Clones and mutants are clearly visible.
Note that with increasing $k$ the classical trajectory becomes
more and more tilted with respect to $Oxy$ plane. } 
\label{quasi}
\end{figure}

    There exist intensive analytical studies devoted to the eigenstates of
$L^2$ and ${\cal L}_3$. When $\eta$ is real they are called intelligent 
spin states  [1] 
and quasi intelligent spin states if $\eta$ is complex  [2]. 
These states extend the well known work of [3]. 
Ref. [2] 
has discussed fully the use of the SU(2) algebra (\ref{lab3}). 
Obviously our WP are not eigenstates of $L^2$  but can
be expanded in such a basis of intelligent or quasi intelligent spin
states with the freedom, by a convenient choice of N, to concentrate the WP
on the sphere. More details on these WP can be found in our recent papers
 [4] and [5].
 
  Let us now sketch briefly the dynamics which take place if one take
these WP as initial WP at time zero and if we let them evolve assuming a
rigid body spectrum. Here we rely fully on the work of Averbukh and
Perelman  [6]. 
For times of the form $t=(m/n)\,T_{\text{rev}}$ $(T_{\text{rev}}=2\pi/\omega_0)$ 
the WP is
subdivided into $q$ fractional WP ($q=n$ if $n$ is odd, $q=n/2$ if $n$ is even),
the shape of these WP depends on the squeezing parameter $\eta$. 
By changing $\eta$
and $k$ one can modify the quantum angular spread and make it different for
the variable $\theta$ and for the variable $\phi$. 
For $\eta=\pm 1$ and for all $m/n$ the
fractional WP are all clones of the initial one (upper part of 
Fig. \ref{ringclon} for
$m/n=1/3$ and $1/4$). For different values their shape changes (we have called
these WP mutants). These shapes are shown in the lower part of 
Fig. \ref{ringclon} and on Fig. \ref{quasi}.
  
  The differences between real and imaginary $\eta$ are not very significant
as shown in Fig. 2. Therefore there exist numerous possibilities for
constructing angular coherent states using these intelligent and quasi
intelligent spin states. Obviously the choice made in (\ref{lab1}) of an 
exponential
WP does not exhaust all possible ones. These remarks illustrate
the richness of the rotation of a rigid body in quantum mechanics. The
internal rotational degree of freedom (i.e.  the use of $D^{I}_{MK}$ functions
instead of $Y^I_M$) can be studied on a similar footing [5]. 

  This work extends to the rigid rotor in three dimensions the revival
mechanism discussed in  [7] 
for the hydrogen atom. The cloning
mechanism, valid in our case only for $\eta=\pm 1$, was already investigated in
ref [8] 
for an infinite square well. A review of the evolution of localized
WP is found in  [9]. 

\begin{center} REFERENCES \end{center}
\begin{description}
\item[{[1]}]  Aragone C et al. 1974 J. Phys.  A : Math.  Nucl.  Gen.  17 L149;\\
  Aragone C et al. 1976 J. Math. Phys. 17 1963.
\item[{[2]}] Rashid M A 1978 J.  Math.  Phys 19 1391, 1397.
\item[{[3]}] Radcliffe J M 1971 J.  Phys.  A4 313.
\item[{[4]}] Arvieu R and Rozmej P 1999 J. Phys. A: Math. Gen. 32 2645.
\item[{[5]}] Rozmej P and Arvieu R 1998 Phys Rev.  A 58 4314.
\item[{[6]}] Averbukh I Sh and Perelman NF 1989 Phys.  Lett.  A 139 449.
\item[{[7]}] Da\u{c}ic-Gaeta Z and Stroud CR 1990 Phys. Rev. A 42 6308.
\item[{[8]}] Aronstein Dl and Stroud CR 1997 Phys. Rev. A 55 4526.
\item[{[9]}] Bluhm R et al.  1996 Am. J. Phys. 64 944.
\end{description}


\end{document}